# Augmented Reality for Maintenance Tasks with ChatGPT for Automated Text-to-Action

Fang Xu, S.M.ASCE,[1]; Tri Nguyen [2]; Jing Du Ph.D., M.ASCE[3]

[1]Ph.D. student, Informatics, Cobots and Intelligent Construction (ICIC) lab, Dept. of Civil and Costal Engineering, Weil Hall 360, University of Florida; Email: xufang@ufl.edu

[1]Undergraduate researcher, Informatics, Cobots and Intelligent Construction (ICIC) lab, Dept. of Civil and Costal Engineering, Weil Hall 360, University of Florida; Email: tri.nguyen@ufl.edu

[3]Associate Professor, Informatics, Cobots and Intelligent Construction (ICIC) Lab, Dept. of Civil and Costal Engineering, 460F Weil Hall, University of Florida, Gainesville, FL 32611 (corresponding author); Email: eric.du@essie.ufl.edu

**ABSTRACT**

Advancements in sensor technology, artificial intelligence (AI), and augmented reality (AR) have unlocked opportunities across various domains. AR and large language models like GPT have witnessed substantial progress and are increasingly being employed in diverse fields. One such promising application is in operations and maintenance (O&M). O&M tasks often involve complex procedures and sequences that can be challenging to memorize and execute correctly, particularly for novices or under high-stress situations. By marrying the advantages of superimposing virtual objects onto the physical world, and generating human-like text using GPT, we can revolutionize O&M operations. This study introduces a system that combines AR, Optical Character Recognition (OCR), and the GPT language model to optimize user performance while offering trustworthy interactions and alleviating workload in O&M tasks. This system provides an interactive virtual environment controlled by the Unity game engine, facilitating a seamless interaction between virtual and physical realities. A case study (N=15) is conducted to illustrate the findings and answer the research questions. The results indicate that users can complete similarly challenging tasks in less time using our proposed AR and AI system. Moreover, the collected data also suggests a reduction in cognitive load and an increase in trust when executing the same operations using the AR and AI system.

## INTRODUCTION

Augmented Reality (AR) has gained a significant attention across diverse sectors including gaming, healthcare, education, and engineering, attributed to the rapid growth of computing powers and advances in hardware for a more efficient environmental understanding and information rendering relevant with the real world (Xiong et al. 2021). Among all successful implementations of AR technologies, AR-supported operation and maintenance (O&M) has shown a great potential in facilities management (Cheng et al. 2020). AR applications in these tasks involve overlaying digital data onto the physical environment and objects, facilitating workers with real-time information to improve task execution and decision-making capacities (Palmarini et al. 2018). Benefits of AR in facilities maintenance tasks have been well documented, including improved spatial awareness (Koch et al. 2014), more integral use of metadata (Palmarini et al. 2018), more accurate risk assessment (Wang and Piao 2019), and enhanced collaboration (Ammari and Hammad 2014).

Despite the cumulative evidence of the benefits of utilizing AR in complex O&M tasks, concerns still persist related to the efficient and automated conversion of context data (e.g., space, environment and tasks) into a format suitable for AR visualizations (Joseph 2019; Vinumol et al. 2013). In this study, we are especially interested in the so-called "text-to-action" challenges in O&M tasks, i.e., a gap between textual information provided as the work instructions and executable actions. In most O&M tasks, instructions are commonly delivered in textual formats, including verbal and written languages (Marocco and Garofolo 2021). On-site workers are required to process these textual instructions, convert them into meaningful semantic meanings, and translate them into specific actions (Du et al. 2020). This process is inherently challenging because textual information is abstract and may lack the specificity required for actions. For instance, an instruction like "tighten the bolt" may require a comprehensive understanding of the text contexts to identify which bolt to operate or in what order. Without the ability to comprehend information in the context of the complete textual instruction, it could lead to ambiguity and errors, particularly in complex operations where precise actions are required. In addition, translating textual information into action requires a high degree of cognitive effort. The worker must not only understand the instruction but also mentally map it onto the physical world, plan the necessary actions, and execute them. Mayer's multimedia learning theory (Mayer 2002) suggests that humans process information via phonological (related to texts) and visuospatial channels, where processing visuospatial information is easier compared to phonological data especially in stressed conditions (Holmes and Bourne 2008). Thus, extracting crucial operational sequence steps from raw textual data to generate comprehensive AR guidance, such as animations, is far from straightforward.

Traditionally, extracting semantic meanings from textual information is mainly based on natural language processing (NPL) tools (Hua et al. 2015). Though effective in many applications, these NLP-based approaches face challenges in supporting the text-to-action needs in complex O&M tasks. Traditional NLP methods are less effective in understanding the context of a large amount of textual data all together (Graves 2018; Yang et al. 2023). O&M tasks require context-sensitive interpretation of instructions, while traditional NLP methods typically rely on fixed rules (e.g., first order logic) (Singh et al. 2020), ineffective to capture these context-dependent meanings when data is oversized. Even with the recent recursive network architecture in NLP, the linear nature of processing textual data makes the computation increasingly expensive when the raw data is big in size, such as processing and understanding the key information from a long paragraph (Vaswani et al. 2017). In addition, O&M instructions can be ambiguous or vague, and workers may use a variety of different expressions to refer to the same action or object. Traditional NLP methods, which tend to rely on exact matches or statistical patterns, can struggle with ambiguity and variability in language (Hassan et al. 2021; Liu et al. 2023).

This paper aims to leverage the recent breakthroughs in Large Language Models (LLMs) to support a real-time and accurate text-to-action generation in AR-based O&M tasks. LLMs, such as OpenAI's ChatGPT, have demonstrated capabilities in filtering and summarizing information, parsing intricate data, and translating this information into actionable steps (Yang et al. 2023). Notable examples of LLMs deployment in comparable contexts abound. For example, ChatGPT has been employed to translate complex legal language into more accessible lay terms (Choi et al. 2023). Another intriguing application includes the use of ChatGPT to produce step-by-step assembly instructions from a high-level description of the goal (Ye et al. 2023; You et al. 2023). These instances hint at the vast possibilities LLMs present in the context of AR for complex O&M tasks. We will show a porotype system we designed to integrate ChatGPT into AR systems to deliver more efficient, accurate, and user-friendly O&M procedures. Preliminary findings from a user test will also be reported to show the performance and cognitive performance with the proposed system. Through a thorough exploration of the interaction between AR and ChatGPT, we aim to showcase the potential of these combined technologies in revolutionizing the O&M tasks.

**LITERATURE REVIEW**

*AR for Maintenance*

In the past two decades, significant efforts have been made to explore the use of AR for operational guidance in O&M tasks within construction and facilities management. It has been found that implementing AR in on-site O&M tasks could substantially increase efficiency and reduce maintenance cost (Palmarini et al. 2018), enhance operational safety (Chen et al. 2020), improve quality and integrity of built assets (Liu and Seipel 2018), and increase customer satisfaction (Liu and Seipel 2018). Successful implementation examples have been widely reported in recent literature such as (Hajirasouli et al. 2022) and (Künz et al. 2022). Nonetheless, AR for O&M applications does not come with no limitations. As one of the earliest attempts to forecast potential and challenges related to AR in AEC/FM disciplines, Chi et al. (2013) performed a comprehensive reviews on more than 100 AR studies at that time, and found that AR had shown the potential to transform practices in many AEC/FM tasks, while challenges related to context-awareness, access to metadata and field information, localization, and portability of AR devices remained to be resolved. Despite the advances in both AR technologies and application workflows, similar challenges have still been reported in recent literature review studies, such as (Kolaei et al. 2022).

Literature has attempted to resolve the issues with AR by incorporating technological advances in other disciplines. Baek et al. (2019) integrated computer vision techniques into AR system that substantially improved the real-time localization of AR for facilities maintenance tasks. Schaub et al. (2022) utilized point cloud data for improved AR localization in FM tasks. Siegele et al. (2020) examined and tested non-vision methods for AR indoor localization, such as Bluetooth, infrared lights, and ultrasonic sensors. Servières et al. (2021) found that the vision and non-vision based data could be merged to support a more robust SLAM analysis for improved localization.

Herbers and König (2019) adopted a data fusion approach to integrate data from BIM, point cloud, and pre-established configuration data (or the "template") to generate a more accurate estimate of the locations. While Corneli et al. (2019) and Chen et al. (2021) leveraged deep learning to improve AR localization when data is sparse.

Another challenge the literature has been striving to revolve relates to the context-understanding of AR in O&M tasks. A popular approach is to incorporate facility metadata (assuming that such data is conveniently available) in AR systems. For example, Chen et al. (2020) developed an architecture to transfer BIM data into AR analytical flow for on-site facilities management tasks. It was found that leveraging the authoring data from BIM could significantly streamline the information retrieval and thus greatly revolve the context awareness issues as identified in previous literature. Similarly, Alavi et al. (2021), Chung et al. (2021), and Machado and Vilela (2020) explored the integration of BIM metadata in AR workflow for various FM tasks, and found that such an integration would not only improve the context-awareness, but also the localization performance of AR applications. Sidani et al. (2021) provides a more throughout review of BIM-AR integration methods and success cases, reinforcing the benefits and importance of ensuring context-aware AR for O&M tasks with metadata from popular authoring tools such as BIM.

However, it should be noted that many O&M tasks and workplaces where they are carried out are dynamic rather than pre-defined. Operational sequence can reorder depending on specific circumstances and system status (Lu et al. 2020). While the workplace, such as the facilities and the surrounding environments can change too because of unexpected scenarios or anomalies (Lu et al. 2020). As a result, O&M workers often need to rely on other hands-on information, such as the data and instructions about the facilities to make spontaneous decisions on the fly (Bortolini and Forcada 2020). It requires a more efficient processing of information, usually in the form of textual data, to come up with coping and operational steps. Relying on pre-established metadata (such as BIM models and point cloud data) and original operation plans would not be sufficient to meet the needs for most real-world O&M tasks. Rather, a certain level of intelligent context understanding would still be needed.

*ChatGPT for Context Understanding*

Traditionally, context understanding has been an effective application domain of Natural Language Processing (NLP). NLP, as a field of AI, allows machines to understand, interpret, and generate human language in a valuable and meaningful way (Chowdhary and Chowdhary 2020). The AEC/FM literature has long recognized the importance of utilizing NLP tools for improving automated context understanding in domain-specific tasks. For instance, Zhang and El-Gohary (2017) leveraged NLP in automated code checking to ensure the building compliance. This method has been extended and improved to support more complex operations such as off-site construction with the help of BIM data (Wong Chong and Zhang 2021). Other successful applications of NLP in AEC/FM tasks include document management (Wu et al. 2020), safety management (Cheng et al. 2020), risk management (Lee et al. 2020), and user requirements analysis (Zhou et al. 2019).

One of the relevant functions of NLP for AEC/FM applications is text summarization, a task aiming to produce a concise and fluent summary while preserving key information and overall meaning (Awasthi et al. 2021). Early works in text summarization adopted extractive approaches, where key sentences or phrases are identified based on certain features and then extracted to form the summary, such as frequency of word occurrence, sentence position in the text, or sentence length (Tas and Kiyani 2007). Later, more robust approaches have been tested to improve the text summarization and comprehension ability of NLP, such as the first order logic (FOL) approach (Zhang and El-Gohary 2017). Recently, recursive networks are more popular in processing complex textural information (Klyuchnikov et al. 2022; Socher et al. 2011). However, the computation can become progressively costly due to the linear nature of processing textual data, especially when dealing with large-sized raw data, like extracting and comprehending key information from an extended paragraph (Vaswani et al. 2017). The recent breakthroughs in large language models (LLMs), especially ChatGPT, provide a promising solution for a more robust context understanding. ChatGPT is a transformer-based language model that is adept at generating contextually accurate text responses given a string of input (Vaswani et al. 2017). The use of such language models in automating information filtering has emerged as a compelling research topic. Although still in the early phase of investigations, promising evidence has been well documented in the recent months about the efficacy of ChatGPT in information filtering. Wei et al. (2023) developed a zero-shot information extraction method with ChatGPT 3.0 and found that it was much more effective than other NLP methods in extracting key information from the unannotated text with little human intervention. Omar et al. (2023) compared ChatGPT 3.0 with other language models from extracting knowledge graphs in Question-Answering systems (QASs) and found that ChatGPT outperformed other models in key metrics such as explainability, question understanding, and robustness. Tan et al. (2023) extended the test to more than 190,000 cases of knowledge-based question answering (KBQA) tasks and found that ChatGPT consistently outperformed other models and achieved human-level correctness. More recently, You et al. (2023) and Ye et al. (2023) tested ChatGPT-4 in facilitating robotic communications with human operators in human-robot collaborative tasks, and found that ChatGPT did not only improve the performance of the task, but also improved human-robot trust. In a more popular domain, ChatGPT has been tested for improving healthcare services, communications, and education (Hopkins et al. 2023; Johnson et al. 2023; Kleesiek et al. 2023).

In a recent report (Bubeck et al. 2023), researchers from Microsoft claimed that the intelligence manifested by ChatGPT was very close to the so-called "*Artificial General Intelligence (AGI)*", i.e., a theoretical type of AI that possesses human-level cognitive abilities to learn, reason, solve problems, and communicate in natural language. The academia is still not entirely clear why ChatGPT has shown such a high-level intelligence, but increasing evidence (Campello de Souza et al. 2023; Holmes et al. 2023; Roazzi ; Törnberg 2023) starts to point to the "*emergent properties*" phenomenon, where new behaviors arises spontaneously and unpredictably as models grow larger, enabling the expansion of its initial capabilities without the need for explicit programming or additional training (O'Connor 1994). Neural science suggests that human intelligence and consciousness could be a result of emergent properties (Hampshire et al. 2012). Given that ChatGPT-4 model possess more than 100 trillion parameters (ABIO 2023), comparably complex as human brains, it may be the foundation for the emergence of its human-level intelligence. We found that the following unique features of ChatGPT can be leveraged to understand complex O&M contexts in the form of textual instructions. First, ChatGPT exhibits a high degree of adaptability and precision in human-level comprehension, primarily due to its training methodology and the underlying technology. As a large-scale language model, it has been trained on a diverse range of data from the internet, which encompasses numerous communication styles and patterns. This vast training base enables it to understand nuanced and contextually appropriate information, crucial for effectively understanding complex meanings. Second, the "smartness" of ChatGPT primarily comes from its transformer-based architecture, more specifically the GPT (Generative Pretrained Transformer) design. Transformers are a type of model architecture used in natural language processing that utilize attention mechanisms to understand the context and relevance of words in a sentence. The model leverages this architecture to understand the context of a conversation, predict appropriate responses, and generate human-like dialogues. Additionally, ChatGPT's design enables it to learn and adapt to new contexts and inputs over time. This is particularly beneficial when digesting and comprehending real-world textural instructions, as these O&M instructions can vary widely.

**SYSTEM DESIGN**

*Overview*

The design of the proposed system prioritizes efficiency, accuracy, and compatibility for real-life applications. There are few challenges to address, first being the integration of three major components: optical character recognition (OCR), the ChatGPT language model, and a virtual environment for managing visuals and interactions. The second challenge involves modeling of the virtual environment, spatial registration, and management of virtual prompts, all of which occur within the Unity game engine. The final challenge lies in the organization and management of text instructions and output commands, enabling the GPT to understand and process the corresponding information, and thus localize problems encountered during operation.

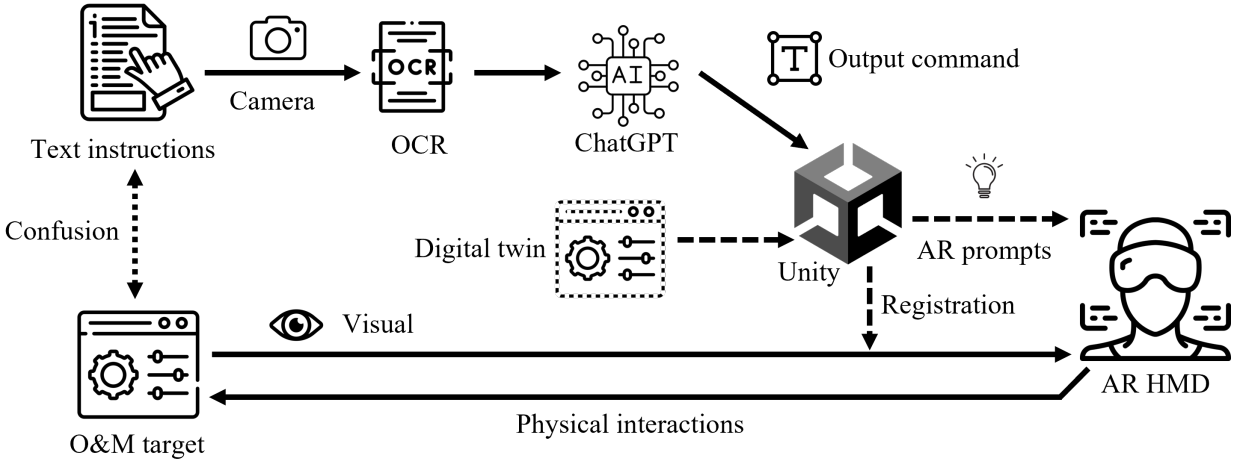

**Fig. 1.** System architecture.

*System Architecture*

The system's architecture, illustrated in Figure 1, positions the Unity game engine (Unity 2023) at the core of the system. Unity forms a bridge between the virtual and physical worlds, facilitating the management of interactions and data exchanges between them. The workflow starts with visual inputs and conversion from 2D images to text content understandable by an AI language model. Images of the instructions could be manually taken by the user and processed with OCR within the same AR system that is connected to online services. The OCR function (Mithe et al. 2013), and the integration of the GPT server are dedicated for processing input text information and generating instruction commands. The Azure Cognitive Services Computer Vision client library (Pterneas 2022) was implemented with Python to enable the OCR function. For an intuitive system design, the front facing camera of HoloLens2 (Ungureanu et al. 2020) was used for taking pictures which was aligned with the user's perspective. The interactions with the camera's shutter and all virtually constructed and managed by the Unity game engine, further discussion of which will follow in the next section. After the image is processed by OCR, text content is received and sent to the ChatGPT online server (Ye et al. 2023), where it requests processing with the GPT-4 model and awaits a response. After receiving a text-based operation command, the strings are sent to Unity using a Python-Unity socket, which facilitates two-way communication between the Python script and Unity application. The network hosts an internal web server with a loopback address (127.0.0.1) on the same machine. In the Python script, a TCP/IP socket server is set up using the "socket" library, while on the Unity side, a TCP/IP client is set up with the "*System.Net.Sockets*" library (Zhou et al. 2020). The connection is established during the execution of both applications, and text data can be sent from the Python script to the Unity application over the socket connection in the form of strings. After receiving the text commands, a Unity function segments the string command and stores them in an array. Each string matches with a unique virtual object in Unity and can be invoked by the user with a click of a virtual button in the AR environment. The visual prompts regarding how to operate the physical buttons then appear on the digital twin model, which is spatially registered on the physical model. The user can then interact with the physical workspace following the AR prompts.

*Virtual Environment and Interactions*

To manage the intricate spatial information and registration of interactable virtual items associated with the physical world, a method for spatial understanding is indispensable. The Mixed Reality Toolkit (MRTK) (Microsoft 2023), provided by Microsoft, offers two specialized features to tackle spatial information processing tasks for HoloLens2: the Azure Spatial Anchor (ASA) (Ong et al. 2021) and World Locking Tool (WLT) (Guan et al. 2023). Leveraging ASA and WLT, HoloLens2 is used in our system to comprehend its surroundings and record spatial information with its spatial awareness system. Coupled with Simultaneous Localization and Mapping (SLAM), HoloLens can also correlate its position with the spatial world in real-time (Ungureanu et al. 2020). Spatial anchors are recognized anchor points, defined by the user, possessing unique 3D features recognizable by HoloLens's spatial understanding function. These identified anchor points are then assigned holograms, thereby stabilizing, and aligning them to the physical world. Consequently, the digital twin of requisite interaction items overlay the physical workspace, aiding user interactions. Given the physical workspace's intricate 3D structure, spatial anchoring methods are implemented instead of visual recognition due to the workspace's unstable appearance and changing illumination in real-life scenarios.

Interactions are managed by Unity, and a virtual scene is preconstructed prior to operation. For instance, the items needed for task design in a case study are displayed below. A box-shaped vacuum device's control unit, with a complex layout of buttons, toggle switches, sockets, and gauges, is shown in Figure 2, alongside its digital model. The digital and physical control units maintain identical measurements and layouts, and the door of the virtual model can be accessed via virtually controlled buttons. At the precise locations of each interactable physical item on the control box, a visualization of operation prompts is attached and can be invoked whenever a command from the AI language model is received.

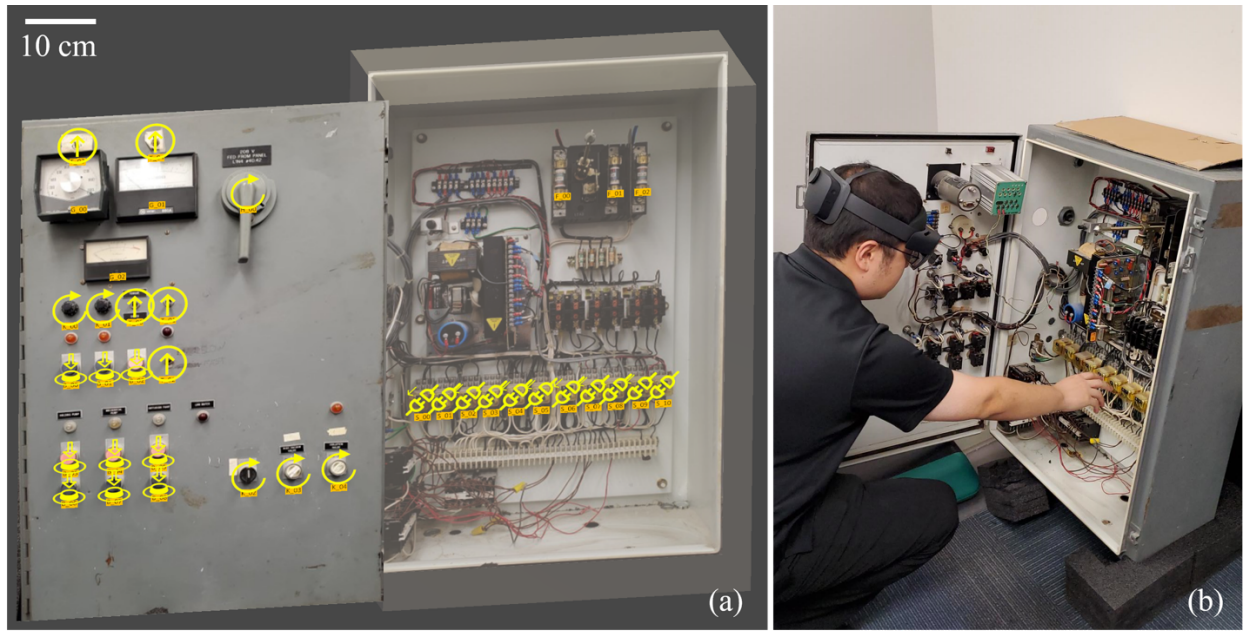

**Fig. 2.** Virtual and physical control panels, (a): the virtual control box with all prompts turned on; (b): a subject interacting with the physical control box with the help from AR HMD.

Interactions managed in Unity derive input from hand interactions using OpenXR and are controlled by MRTK's command system (Takayuki 2020). Hand interaction in AR provides the user with intuitive manipulation capabilities of virtual objects and command accessibility without exiting the virtual scene or relying on any physical input systems. A menu attached to the user's hand manages all interaction items with virtual buttons, which can be pressed directly with the hand. The hand menu can be summoned when the user places their palm against their face while maintaining a flat palm posture, and it can be hidden simply by moving the hand away. The expandability of virtual interactions also implies that the system can be readily integrated or deployed with any desired operations and can easily accommodate tasks of varying complexity.

*Integration of ChatGPT and Prompt Design*

To ensure that the returned commands from the language model are proper and accurate, a specific input of text instructions for interpreting a certain type of content is required. By understanding the construction of the GPT model, as well as running a series of pilot studies, we found that the text context with complex logic loops was not ideal for the ChatGPT model to understand, while lengthy but sequenced text can be easily filtered and summarized into step-by-step operation commands with a good quality. As a result, we focus on long textual instructions with linear logic steps, i.e., no recursive, embedded or nested logic. Figure 3 shows an example of how ChatGPT successfully convert a long wordy instruction into simply sequence orders, and how AR prompts are related with the instructions.

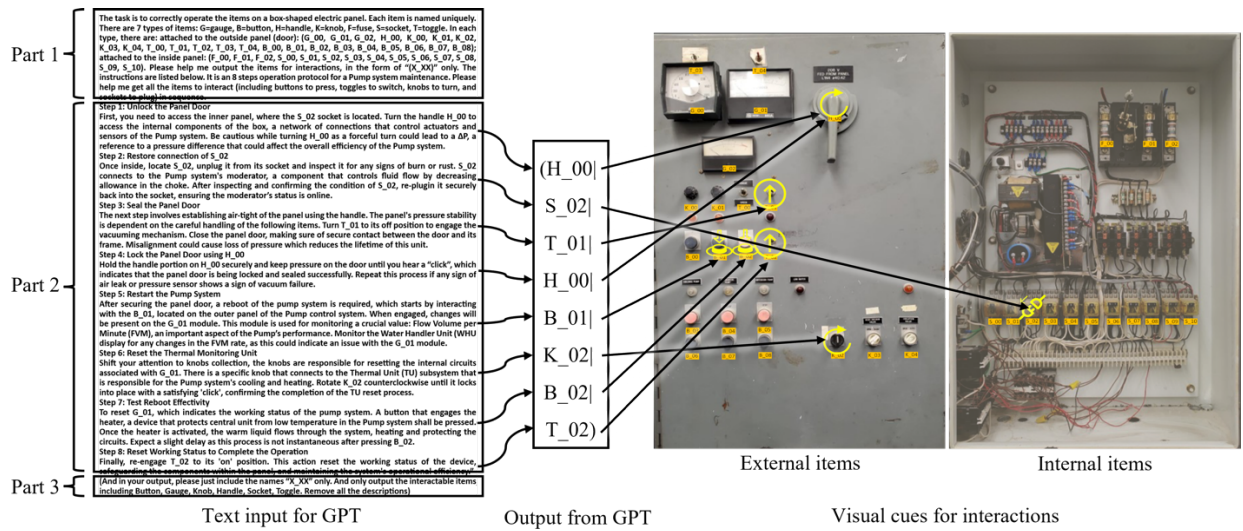

**Fig. 3.** Instruction interpretation and prompt management.

Specifically, we propose that the prompt should is composed in three modulated parts to help GhatGPT get a clear understanding of the context and output the correct commands for Unity to handle. The proposed prompt design is shown in Figure 4.

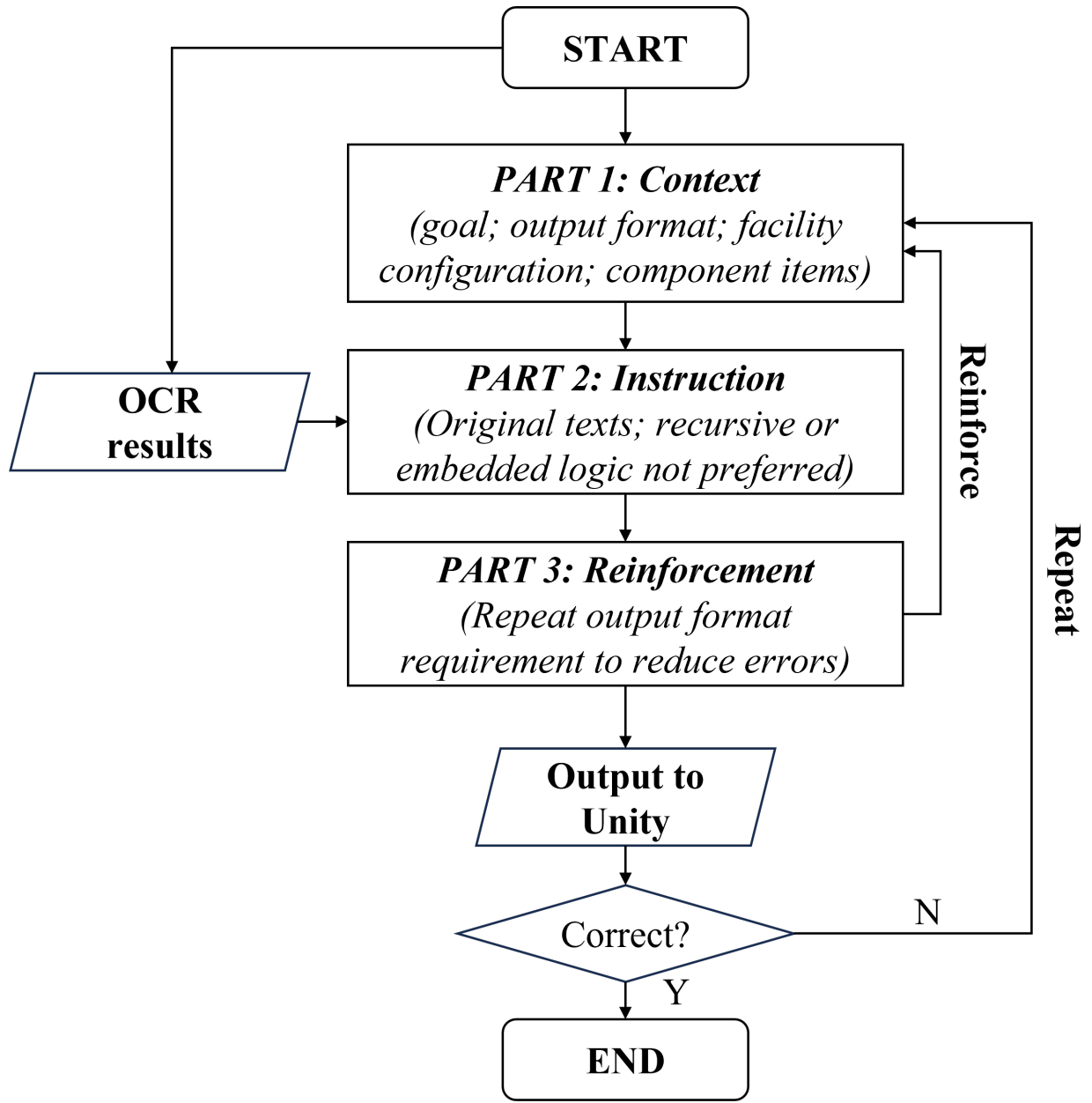

**Fig. 4.** Prompt design.

As shown in Figure 4, the first part of the prompt is context, i.e., a general description of the operation as well as the system, which introduces all the interaction operations and the context of the operation, as well as listing the names of all the interactable items. The second part is the instruction which includes the raw texts for operation instructions. The third part reinforcement repeats instructions for GhatGPT to output the desired format of commands. It is worth noting that with this three-part instruction layout, if the second part matches the items within the context of the first part instruction, it can be replaced with any context and the user can still get correct output from GhatGPT. This design will ensure the flexibility of input commands and the stability of output prompts. Figure 5 shows an example of the proposed workflow inside the API environment. After a picture is taken and sent to OCR, the output text can be sent to ChatGPT and awaiting response. Upon receiving output text from ChatGPT, the message can be decomposed and stored in an array for Unity to manage. A short demo of the system can be found here: https://youtu.be/ETOyMTe8zmA

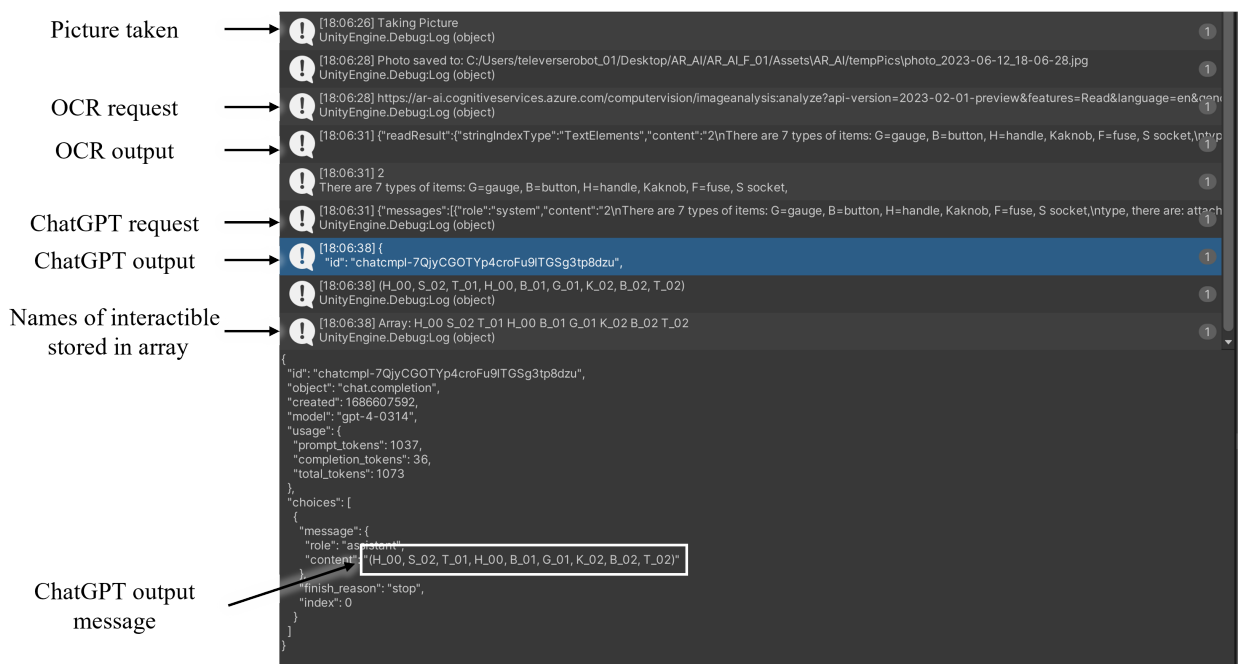

**Fig. 5.** Example GhatGPT output revealed in Unity editor from the proposed system.

The proposed AR & AI system showcases intuitive hand interactions for manipulating both virtual and physical objects, as well as obtaining operation cues. The entire virtual environment is managed by Unity game engine, and with the aid of an AR head-mounted display (HMD), visualizations and interactions can be seamlessly achieved between the virtual and physical worlds. The system's ability to efficiently process instructions and assign prompts to actual interactable objects may potentially enhance user performance in real-life operations and maintenance (O&M) tasks. The subsequent section presents a case study that supports this argument.

**EXPERIMENT**

A case study (n=15) of the proposed system was conducted to testify the system. The experiment was conducted with 15 subjects recruited publicly aged from 18 to 30 years old with an average of 24.8. Three out of 15 subjects reported to be experienced with AR HMDs, and hand interactions. Two out of 15 subjects claimed to have little motion sickness when using the HoloLens2, and the rest 13 subjects didn't feel any sign of sickness or dizziness during the experiment. The experiments were performed using Open AI key version: "gpt-4".

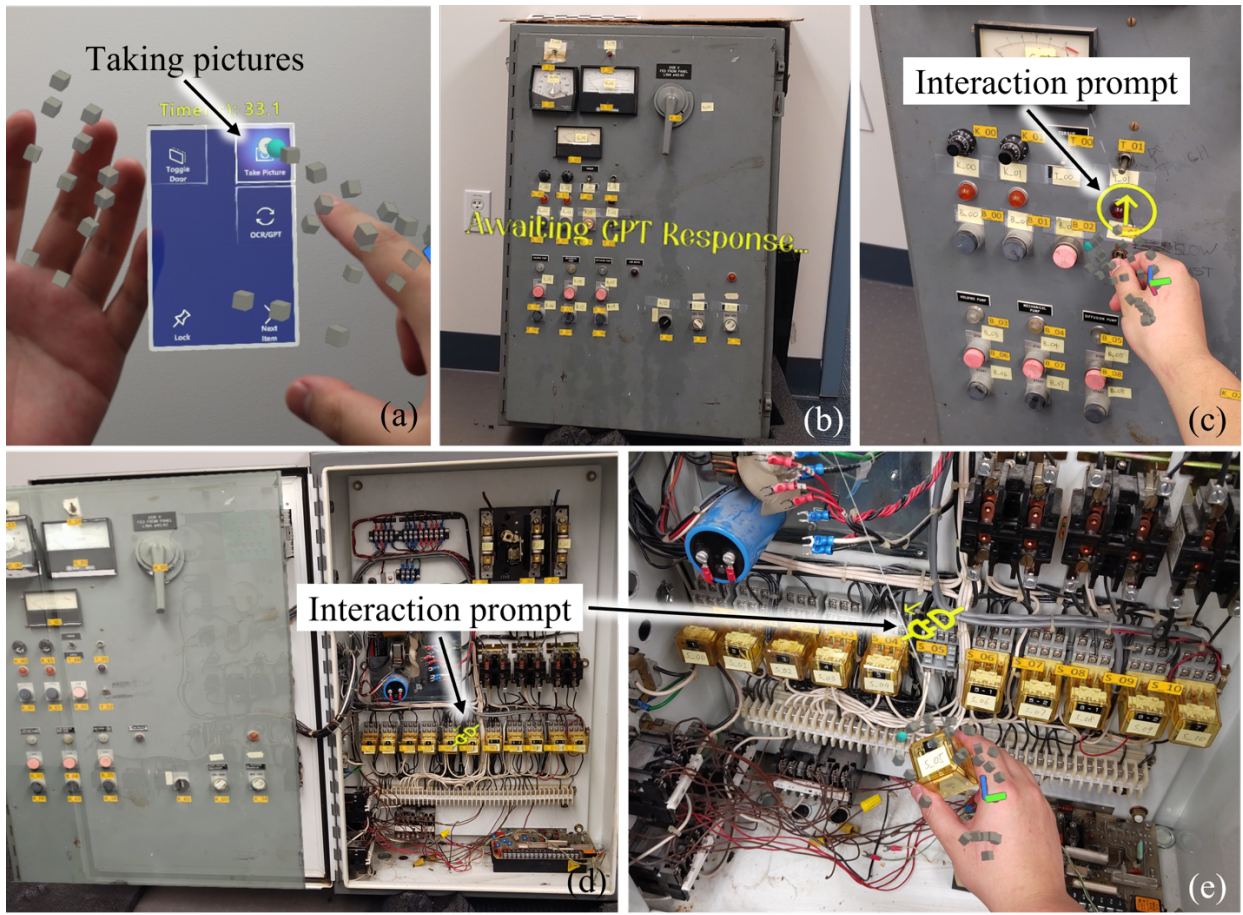

**Fig. 6.** Workflow and interactions, (a): taking a picture with virtual camera shutter; (b): waiting for ChatGPT output; (c): interacting with a toggle button; (d): accessing internal items; (e): unplugging a socket according to the AR prompt.

A box-shaped control panel from a vacuum system was used to represent the physical workspace, and a digital twin model of the same box was assembled virtually in a Unity scene (Figure 2). The interactions with the AR virtual scene are made with MRTK 2.8 and was enabled with hand interactions input. A hand menu (Figure 6 (a)) controls all the interactions that manage the operation in the virtual space. The physical environment surrounding the box was scanned with HoloLens2 prior to the operation, and the digital model was anchored at the same physical position of the box. There were two layers of interactable items attached to the box, which was the outer panel and the internal components. On the outside panel, there were three types of interactable items, which were: gauge (G), button (B), handle (H), knob (K), and toggle (T), while on the inside, there were fuse (F) and socket (S). The names of the interactable objects were G_00-02, H_00, K_00-04, T_00-04, B_00-08, F_00-02, S_00-10. To relate the instructions with the physical items on the box controller panel. The name of each button was clearly defined for Unity to call the items when received commands from ChatGPT. The name of the interactable items was unique and consistent in both Unity editor and text instructions, to make sure of precise interpretations. It is also worth mentioning that the text output from ChatGPT was directly fed to Unity editor without being processed by string searching algorithms, and the output could stay precise throughout the experiment.

The design of this within-subject experiment included two conditions: "no augmentation" and "AR & GPT". The text content for these conditions varied in terms of operation sequence, object names, and scenarios to mitigate the learning effect. In the "no augmentation" condition, subjects would need to troubleshoot a malfunctioning Heating, Ventilation, and Air Conditioning (HVAC) system, with eight operation steps and eight interactive items, comprising a total of 491 words of instructions. Subjects were required to wear the AR Head-Mounted Display (HMD) and interact with the physical buttons in the correct sequence with minimal time lapse. The "AR & GPT" condition also involved interactions with eight items on the control panel and eight instruction steps, totaling 631 words (116 + 489 + 26), for the maintenance of a pump system. The workflow of "AR & GPT" condition is shown in Figure 6. In this condition, subjects were first asked to interact with a virtual "hand menu" to photograph the instruction, which was then processed and uploaded to ChatGPT. Subjects needed to wait approximately 5 seconds for the processed result to return to Unity. In practice, after uploading the instructions to ChatGPT server, the subject was asked to wait until the "Awaiting GPT Response" prompt to disappear for continuing the experiment. When the result was ready, subjects were asked to click the "next" button on their hand menu to navigate through all the items for interaction. Upon clicking "next", a visual prompt appears on the corresponding physical button on the control panel. Since the AR scene was only for cueing the sequence and location, all physical items should be interacted for a correct result. In the digital control box, the external panel could also be toggled open/close with a button on the "hand menu". Figure 6 (c) through Figure 6 (e) shows two interaction examples with the physical buttons, including one toggle button to be flipped and a socket to be unplugged from the control box.

In both conditions, the required interactions occurred in sequence, and they all involved accessing one of the internal components on the control panel. The correct interaction in both conditions included pressing two buttons, turning two knobs, detaching one socket, flipping two toggle switches, and turning the door handle twice. The correct operation sequence for the "no augmentation" condition was: "B_04, K_03, B_07, H_00, S_04, T_04, H_00, T_04", and the correct sequence for the "AR & GPT" condition was: "H_00, S_02, T_01, H_00, B_01, K_02, B_02, T_02". Therefore, the complexity of operation was equivalent for both conditions. Throughout the experiment, all participants were able to correctly operate the virtual buttons, and the system functioned successfully in every trial. Upon completion of each of the two conditions, participants were required to complete two questionnaires, namely the NASA Task Load Index (NASA-TLX) (Hart 2006) and a trust evaluation (Merritt 2011). Both conditions were untimed, and performance was evaluated based on task completion time and the accuracy of interaction sequence.

Prior to initiating the experiment, all participants went over a training session to ensure familiarity with the AR system and the physical control panel. From previous experiences, it was determined that mastering hand interactions in AR could be challenging for inexperienced users; thus, training was essential to prevent technical factors from impacting the results. The training session began with the calibration of the HoloLens2 to ensure accurate input command detection. Subsequently, participants were asked to engage with the "MRTK Example Hub" app provided by Microsoft, which included hand interaction examples to acclimate participants with hologram interactions. Participants were also encouraged to try the actual hand menu used in the experiment to ensure precise button pressing. They were then introduced to the physical control panel, including both external and internal panels. All participants were able to operate the AR virtual environment without experiencing motion sickness or any other discomfort that could potentially affect data collection.

The physical interaction sequence was recorded and documented using a video camera during the experiment, and the completion time was derived from the video stream. For the "no augmentation" condition, the experiment commenced with the participant reading the instruction and concluded with the last item interaction. For the "AR & GPT" condition, the starting point was marked from taking the picture for OCR.

**RESULTS**

In this study, a within-subject factor design was applied with two conditions. The results of each condition were analyzed using the Wilcoxon signed-rank test to determine the relationship between groups, with a null hypothesis criteria p of 0.05. All participants were able to successfully complete the tasks during the experiment, with the longest completion time being 359 seconds. The output from ChatGPT was also documented and compared with the correct operation sequence.

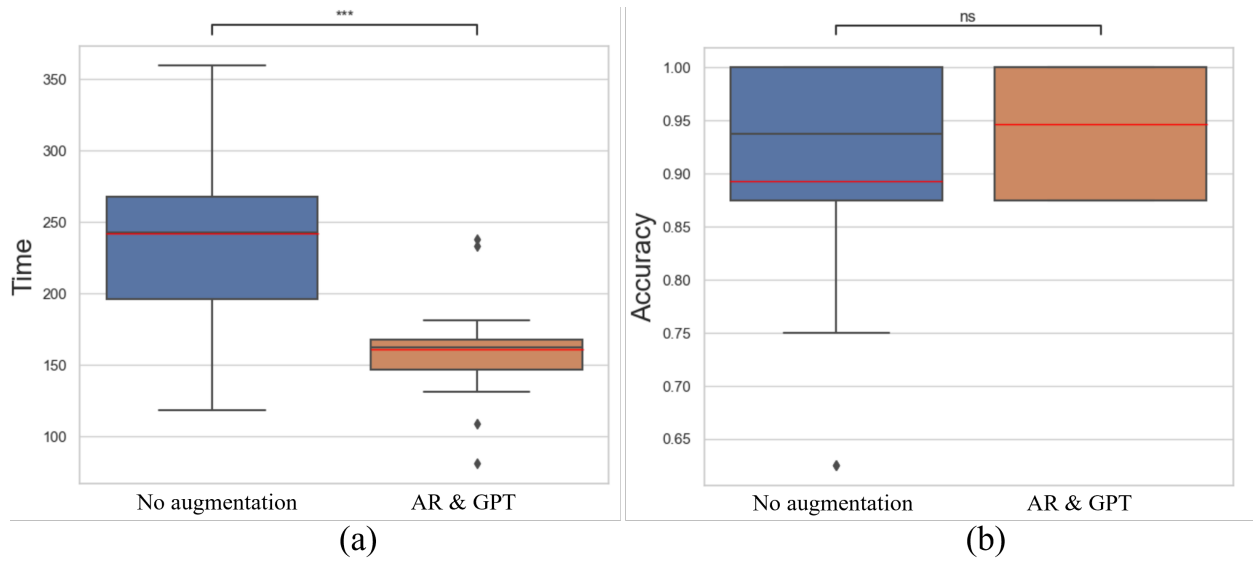

**Fig. 7.** Result of the performance, (a): time of completion; (b): accuracy of physical interactions.

**Table 1.** Performance results in two experiment conditions.

| **Condition:** | **Time of Completion** | | **Accuracy** | |
| --- | --- | --- | --- | --- |
| | Average (s) | STD | Average (s) | STD |
| No augmentation | 241.7 | 72.01 | 0.8929 | 0.1374 |
| AR & GPT | 160.6 | 41.14 | 0.9464 | 0.0642 |

Performance was evaluated by the time of completion and interaction accuracy. For completion time, the "AR & GPT" condition showed a lower average completion time of 241.7 seconds than 160.6 second from the "no augmentation" condition ($p = 0.0006104$) (Figure 7 (a)). The accuracy of physical interaction of "no augmentation" condition resulted in an average of 0.8929 out of 1 (Table 1), which is lower than 0.9464 from the "AR & GPT" condition (Figure 7 (b)). But no significant difference was observed in the accuracy difference between the two conditions, which was caused by the simplicity of tasks. The operation procedure was made obvious for both human eyes and ChatGPT, so there was not a significant difference in the accuracy result, but the significant difference in completion time proved the efficiency of applying such a system to help in O&M tasks. It is worth mentioning that the accuracy of ChatGPT in processing the text content was 0.9553 in average, which was slightly higher than the physical interaction accuracy. This was caused by the registration drift during some of the trials, but most of the users found that AR was helping in the process. In addition, there was 25 seconds waiting time for processing the data with online servers included in the completion time calculation, so the actual action time in the "AR & GPT" group was nearly half of the time consumed in the "no augmentation" condition.

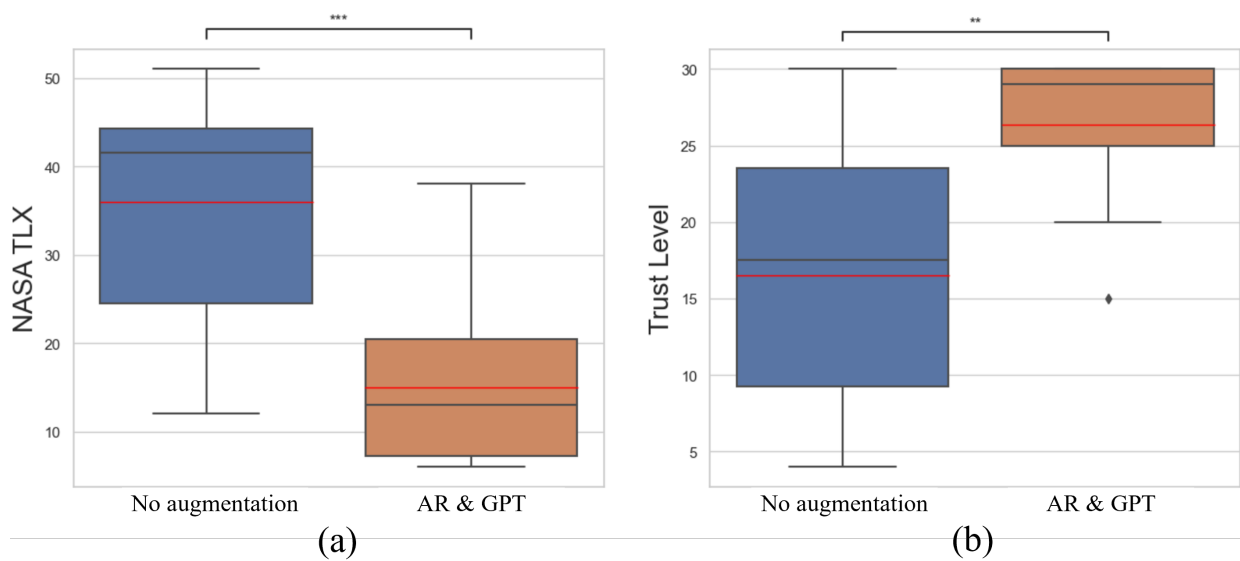

**Fig. 8.** Result of the survey, (a): NASA-TLX; (b): Level of Trust.

And the effectiveness of using AR for helping the subjects relate virtual prompts directly to the physical objects were revealed not only in the performance but also in the survey results. From the NASA-TLX survey, there was a significant difference between the two conditions ($p = 0.000845$) (Figure 8 (a)). This proved that the "AR & GPT" method was successful in relieving the cognitive load during this O&M operation. On the other hand, there was also a significant difference observed in the Trust survey result (p = 0.002509) (Figure 8 (b)), which indicated that while performing the tasks, subjects were able to trust the virtual prompts provided by ChatGPT and AR for helping them finishing the correct interactions.

**DISCUSSION**

The findings from our human-subject experiment (N=15) indicate that ChatGPT-enabled AR outperforms conventional AR methods across several key metrics. This has crucial implications for how we approach complex maintenance tasks and potentially other fields where AR is applied. Completion time is a critical factor in maintenance tasks, where efficiency often translates directly into cost savings. In our experiment, the usage of ChatGPT-enabled AR was associated with quicker task completion compared to the conventional AR methods. This suggests that the integration of the ChatGPT model enhanced the efficacy of AR, allowing users to complete their tasks more swiftly. The precise mechanisms through which this acceleration occurred would be an interesting avenue for future research, although one might speculate that the intuitive, contextually aware guidance provided by the ChatGPT model played a role. The advantage of ChatGPT-enabled AR was also observed in terms of sequence correctness. Sequence correctness in a maintenance task is pivotal as incorrect or sub-optimal sequences can lead to errors, rework, or even equipment damage. The observed superiority of the ChatGPT-enabled AR condition suggests that the contextual understanding and advanced prediction capabilities of the GPT model may facilitate better sequence planning and execution. It underlines the potential of AI to enhance the precision and reliability of AR-guided tasks. Participants using ChatGPT-enabled AR also reported higher perceived trust. Trust in the tool or system used is a crucial element of user acceptance and continued use. The enhanced trust could be attributed to the quality of interaction and guidance provided by the GPT model, reinforcing that integrating advanced AI models with AR can improve user confidence and satisfaction. Lastly, the findings concerning cognitive load, as measured by the NASA Task Load Index (TLX), reinforce the practical benefits of integrating GPT models with AR. Lower cognitive load indicates a more effortless interaction with the system and can lead to better task performance and user experience. These findings align with the underlying premise of AR - to seamlessly blend digital guidance with the physical world, reducing the cognitive effort required to interpret and act upon information.

The following limitations are also identified to be addressed in the future. First, the generalizability of our findings may be influenced by the nature of the tasks used in our experiment. We focused on complex maintenance tasks, which involve a particular set of skills and cognitive processes. It remains to be seen whether the advantages of ChatGPT-enabled AR observed in this context would extend to tasks of a different nature, such as those that are more creative, decision-intensive, or require interpersonal interaction. Second, while our study found that ChatGPT-enabled AR was associated with lower cognitive load and higher perceived trust, these outcomes were based on self-reported measures, which may be subject to bias. For instance, participants' perceptions of cognitive load and trust might be influenced by their prior expectations, their familiarity with AR or AI, or other individual factors. Future studies could incorporate more objective measures of these outcomes, such as physiological measures of cognitive load or behavioral indicators of trust. Lastly, the implementation of ChatGPT in an AR setting is not without challenges. There may be technical constraints or complexities related to integrating these technologies, and these challenges were not explored in our study. Future research should also investigate the feasibility and best practices for implementing ChatGPT-enabled AR in different contexts and scales.

**CONCLUSIONS**

This study explored a novel method of leveraging ChatGPT to streamline and optimize the AR-based O&M tasks. A pressing challenge in O&M tasks is to enhance the efficiency and accuracy of while minimizing cognitive load in digesting complex text-based instructions. In response to this challenge, we proposed the integration of ChatGPT-4 into an AR-based operation system, aiming to harness the power of AI in providing real-time, contextually aware guidance. In our proposed system, when a user interacts with the AR interface - for instance, by watching text instructions - this input is processed by the ChatGPT model. Leveraging its ability to understand language and conduct a reasoning process, the ChatGPT model formulates the proper operational steps, which are then presented to the user through the AR interface. This could take the form of virtual symbols superimposed onto the relevant parts of the environment and facilities. By integrating ChatGPT into the AR system in this way, we aimed to create a more intelligent, contextually aware, and user-friendly AR experience. Rather than simply displaying pre-programmed instructions, the ChatGPT-enabled AR system can provide real-time guidance that adapts to the user's inputs, the state of the equipment, and potentially even the user's behavior.

To evaluate the effectiveness of this approach, we conducted a human-subject experiment involving a complex maintenance task. Participants were asked to perform the task under two conditions: using our novel ChatGPT-enabled AR system and using a conventional AR system. Our findings revealed a significant advantage of ChatGPT-enabled AR over conventional AR methods. The use of GPT-enabled AR led to shorter task completion time, higher sequence correctness, increased perceived trust, and reduced cognitive load. These results indicate the substantial potential of integrating advanced AI models like GPT into AR systems, thus advancing our understanding of the interplay between AI and AR.

From a practical perspective, these findings could have significant implications for various sectors that rely on complex maintenance tasks. The enhanced efficiency, accuracy, and user experience associated with ChatGPT-enabled AR could translate into cost savings, improved equipment longevity, and higher worker satisfaction. This paves the way for the broader adoption of AI-enhanced AR in real-world settings, bringing the benefits of this advanced technology to a wider audience. Our work represents an exciting step forward in the confluence of AR and AI. By demonstrating the potential of ChatGPT-enabled AR to enhance complex maintenance tasks, we hope to inspire further research and innovation in this field, ultimately contributing to a future where technology seamlessly augments human capability, improves task efficiency, and enriches user experience. Further studies are needed to understand the specific features of the GPT model that contribute to improved AR performance, and to validate the applicability of our findings across different tasks and domains. Furthermore, in-depth investigation into the technical feasibility of GPT-AR integration and the identification of best practices in its implementation will be essential for this technology's successful translation into practice.

**DATA AVAILABILITY STATEMENT**

All data, models, or code generated or used during the study are available from the corresponding author by request.

**ACKNOWLEDGEMENTS**

This material is supported by the National Institute of Standards and Technology (NIST) under grant 70NANB21H045. Any opinions, findings, conclusions, or recommendations expressed in this article are those of the authors and do not reflect the views of NIST.